**Harnessing Evanescent Wave Interaction for Enhanced Optical $NO_2$ Detection with Carbon Nanotube-Coated Side-Polished Fiber**

*Egor O. Zhermolenko, Khasan A. Akhmadiev, Aram A. Mkrtchyan, Fedor S. Fedorov, Anastasiia S. Netrusova, Aliya R. Vildanova, Dmitry V. Krasnikov, Albert G. Nasibulin and Yuriy G. Gladush**

Skolkovo Institute of Science and Technology, Moscow, 121205, Russia
E-mail: y.gladush@skoltech.ru

Funding: Russian Science Foundation (Grant No. 22-13-00436 (П)).

Evanescent-wave photonic sensors employing nanomaterial-coated waveguides are commonly interpreted in terms of absorption modulation of the sensing layer. Here, we demonstrate that, in such systems, gas-induced reshaping of the guided optical mode can dominate the transduction mechanism and even reverse the sign of the optical response. Using side-polished optical fibers covered with single-walled carbon nanotube (SWCNT) thin films, we observe a pronounced polarization- and thickness-dependent response to $NO_2$ exposure. For transverse-electric polarization, the optical response decreases with increasing film thickness and changes sign for thicker coatings, whereas for transverse-magnetic polarization, the response increases monotonically. Numerical modeling reveals that modulation of the SWCNT complex refractive index redistributes the evanescent field, alters the mode-matter overlap integral, and controls propagation loss. These results demonstrate that evanescent-wave sensor behavior is governed not only by intrinsic material sensitivity but also by mode reshaping induced by nanomaterial coatings. The identified mechanism provides a general framework for designing fiber-optic and integrated photonic sensors based on evanescent-field interactions with tunable nanomaterials.

## 1. Introduction

Analysis of the environment is expected to be boosted by using miniature sensors, such as pellistors[1], semiconductor gas sensors[2], and electrochemical devices[3], enabling monitoring with lower power consumption to facilitate personal use[4] and integration in Internet-of-Things systems.[5] One of the known pollutants is nitrogen dioxide, $NO_2$, an oxidizing gas that facilitates the formation of $HNO_2$ and $HNO_3$ when absorbed by water, thus, making it also a corrosive agent. $NO_2$ concentration in a workspace, Threshold Limit Value – Time-Weighted Average is restricted to 500 ppb according to The European Commission recommendation[6] while annual average concentration is reported to be 5.2 ppb by the WHO.[7] Detection of this hazardous and corrosive analyte requires using robust and stable sensors.

Gas sensors convert chemical information into analytically useful signals through interactions between target analytes and, often, a functional material.[8] This material serves as the sensing element, where adsorption of volatile compounds alters its physicochemical properties to enable detection. Among material platforms, carbon nanomaterials—particularly graphene and carbon nanotubes—offer distinct advantages, such as ultra-high surface-to-volume ratios[9], maximizing analyte interaction sites, tunable electronic properties[10] that amplify signal transduction, and also mechanical/chemical robustness[11] suitable for harsh environments, their conductivity favors low-noise, resulting in high signal-to-noise ratio, which is anticipated to yield low limit of detection (LOD). These attributes explain their advantage in next-generation sensor designs, overcoming limitations of traditional metal oxides[12], which require high operating temperatures.[13–16] Resistive sensors based on graphene are well-established[17] and exhibit high sensitivity - even enabling single-molecule detection in some configurations.[18] Single-walled carbon nanotubes (SWCNTs) have also garnered significant attention as resistive gas sensors.[19,20] A notable example includes an $NO_2$ sensor using surfactant-free, aerosol-synthesized SWCNT films, achieving a sensitivity of 41.6% towards 500 ppb in the mixture with air, response and recovery times of 14.2/120.8 seconds, and a detection limit as low as 0.161 ppb at 150°C.[21] However, for certain applications, optical sensors offer distinct advantages. These include remote sensing capabilities with free space or fiber-assisted light delivery[22], low power consumption[23], and inherent immunity to electromagnetic interference.[24] Optical sensors can also be utilized in highly flammable environments[25], where temperature-activated sensors might be inoperable. While free space optical measurement typically requires long interaction path lengths to achieve comparable detection limits for volatile compounds[26], application of nanomaterials can reduce the sensing path length down to micron- or even nanometer-scale.[27]

In recent decades, optical fibers have revolutionized communication technology by serving as crucial optical waveguides, paving the way for fiber-based sensors.[28–34] Advances in microfabrication and the emergence of two-dimensional materials have facilitated the integration of nanomaterials onto optical fibers. By removing part of the fiber cladding, a D-shaped cross-section is created in side-polished fibers (SPFs), providing a flat surface for integration of micro- and nanostructures. This configuration favors effective light-matter interactions while maintaining fiber integrity and minimizing the losses. Evanescent wave fiber sensors have demonstrated significant potential for chemical detection, while displaying the diversity of sensitive material platforms (chalcogenides[35], graphene oxide[36], 2D composites[37]), target analytes (VOCs[38], biomolecules[39]). For example, a mid-infrared evanescent wave sensor based on side-polished chalcogenide fibers successfully detected VOCs.[40] The application of graphene oxide coatings on SPFs significantly enhanced sensitivity, achieving three to five times greater sensitivity to butane when compared to unmodified fibers, alongside rapid response times (80 seconds) and excellent repeatability.[40] Similarly, in methane sensing, a mid-infrared fiber sensor employing evanescent wave spectroscopy demonstrated its capability for gas and liquid analyses across a wavelength range of 2515–3735 nm, achieving a detection limit of 0.52% for methane samples.[41] Quartz-enhanced photoacoustic spectroscopy has emerged as a highly sensitive technique for trace gas detection, yet conventional systems rely on free-space optical alignment, limiting robustness in field applications. A significant advancement was demonstrated by Twomey et al.[42], in which a fully integrated side-polished fiber-based sensor was developed. Applied for methane detection, the system achieved a 34 ppmv detection limit at 300 msec integration time, enabling operation in harsh environments and mobile platforms. Notably for biosensing, a D-shaped fiber surface plasmon resonance sensor with a $MoS_2$-graphene composite was developed for glucose detection, achieving a sensitivity of 6708.87 nm $RIU^{-1}$ with enhanced selectivity and stability through functionalization with pyrene-1-boronic acid.[43] In a similar approach for $NO_2$ sensing, an SPF was coated with a two-dimensional plasmonic tungsten oxide to create a strong light-matter interaction at 1550 nm, enabling detection as low as 8 parts per billion.[29] Research has also shown the viability of SWCNTs for optical gas sensing, such as a fiber loop ring-down sensor for detecting $SF_6$ decomposition components, which achieved a sensitivity of 0.183 ns $ppm^{-1}$ and a detection limit of 19.951 ppm for carbon monoxide (CO).[44] In most studies, the attention is concentrated on the sensing capabilities of the covering material, while D-shaped fiber platform is mostly treated as a convenient platform for light delivery and providing required interaction length. The effects of the mode reshaping due to broken symmetry of material-covered waveguide on a sensor performance usually remain outside of research focus.

In this study, we develop an evanescent-wave $NO_2$ sensor by integrating SPFs with SWCNT films and elucidate the underlying interaction mechanisms affecting system response. We demonstrate that the sensors' response cannot be accurately described solely by gas-induced changes in SWCNT absorption; it also requires accounting for alterations in the fiber mode shape due to variations in the SWCNT refractive index induced by gas. This effect can be so pronounced that it reverses the response sign from negative to positive. Through combined experimental and numerical analyses, we optimize sensor parameters to achieve high sensitivity. Additionally, we evaluate performance at elevated temperatures and in humid conditions, confirming applicability to real-world scenarios.

## 2. Results and Discussion

Application in optical sensing demands a SWCNT film with a large surface-to-volume ratio of CNTs (i.e., low bundling), clean SWCNT surface (i.e., absence of surfactant traces and other contaminations), and the ability to control carbon nanotube diameter (i.e., to tune the electronic transitions to the resonance with the chosen probing electromagnetic wave).[45] To meet these requirements, we used the floating catalyst chemical vapor deposition (FCCVD) method to synthesize a series of SWCNT films with various thicknesses, their mean diameter was tuned to have the $S_{11}$ transition at 1550 nm. In this method, carbon nanotubes grow in the gas flow on a catalyst nanoparticle from carbon monoxide disproportionation, which served as the carbon source (**Figure 1A**). A proper tuning of synthesis parameters allows the growth of SWCNTs of high purity with controlled mean diameter (temperature, $CO_2$ content[46]) and bundling degree (ferrocene concentration[46,47]). The aerosol of SWCNTs is collected at the outlet of reactor on a nitrocellulose filter (HAWP, Merck Milipore), where the film thickness can be controlled with high precision from a sub-percolating network to hundred-nanometer-thick films.[48,49] These films are transferred to the surface of the side-polished fibers by a dry-transfer method[50,51], leaving the surfaces of the SWCNTs clean from any surfactant and intact from any contamination. The details of the SWCNT films characterization (SEM) can be found in Supporting Information. Scheme of the optical gas sensor and the working principle are illustrated in **Figures 1B-C**. In this study, a side-polished optical fiber, embedded in a quartz block, served as the sensor platform (**Figure 1B**). The evanescent tail of the propagating mode reaches the polished surface enabling chemical sensing by facilitating interactions between the wave and the surrounding environment (**Figure 1C**). When SWCNTs are deposited on the polished surface, they absorb light at interband transition frequencies, leading to a decrease in transmittance through the fiber. The absorbance spectrum of the films with $S_{11}$ transition maximum at 1550 nm is shown in **Figure 1D.**

In this study, nitrogen dioxide, a strong oxidizing agent, was chosen as the target analyte. Adsorption of $NO_2$ on SWCNTs induces additional *p*-type doping, shifting the Fermi level closer to the valence band. This process decreases the absorbance of the SWCNTs, thereby increasing the light transmittance through the SPF. A schematic of the zone structure and density of states in a semiconducting SWCNT is presented in **Figure 1E**. Simultaneously, the electrical resistance of the SWCNT film was monitored to evaluate its response to $NO_2$ exposure. The resistive sensor served as a reference and provided complementary information, monitoring interactions between $NO_2$ molecules and SWCNTs.

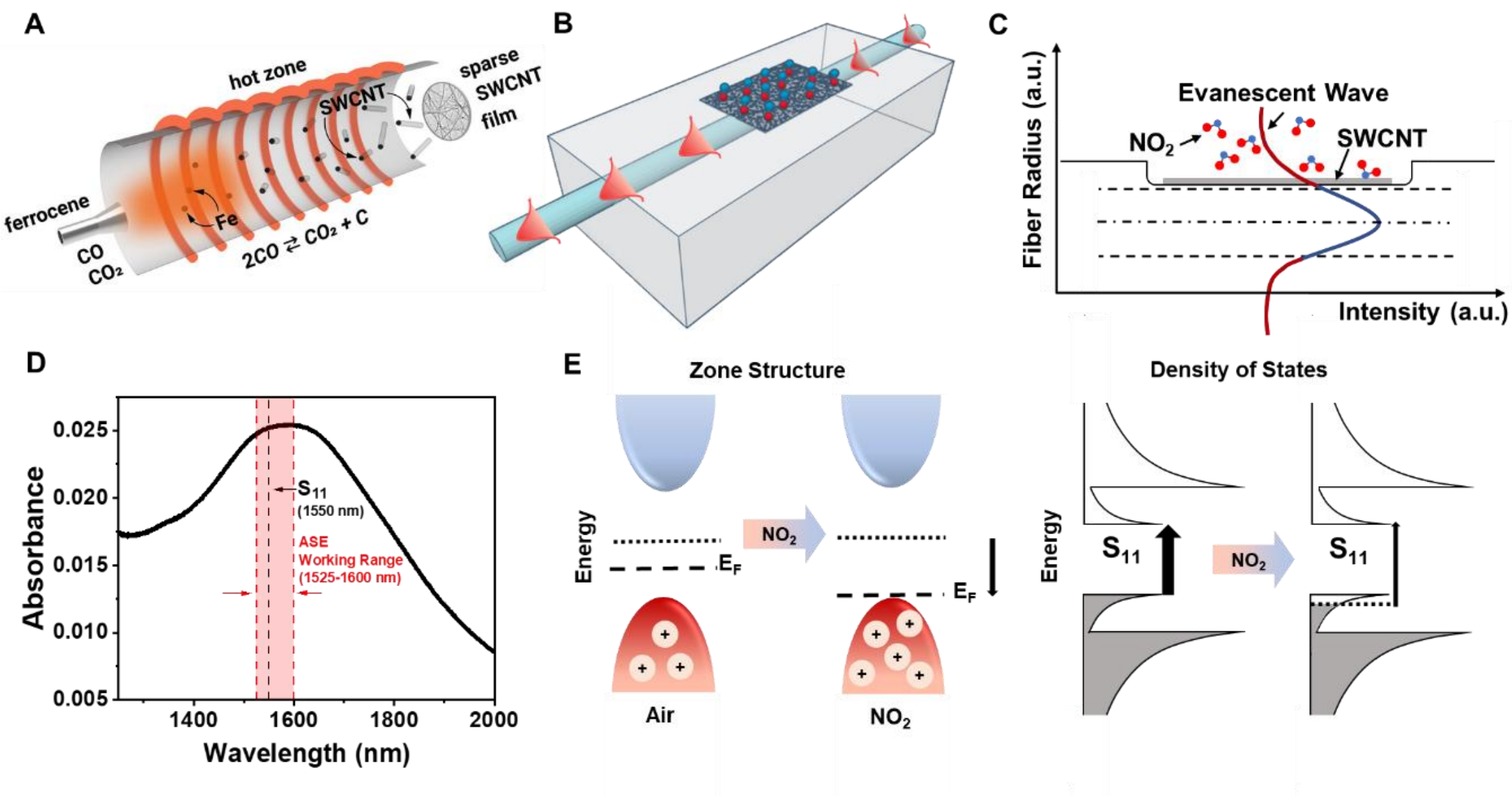

**Figure 1.** A) Aerosol CVD synthesis of SWCNTs. B) D-Shape embedded in a quartz block with SWCNT film deposited on the polished region and adsorbed $NO_2$ molecules. C) Illustration of the interaction mechanisms between analyte, evanescent field and SWCNT placed on SPF. D) Absorption spectrum of SWCNT film. E) Schematic zone structure and density of states in semiconducting SWCNTs.

The experimental setup is depicted in **Figures 2A** and **2B**. An amplified spontaneous emission (ASE) source emitting radiation at 1525-1600 nm was employed as the light source. The optical signal generated by the ASE source was transmitted through the SPF with SWCNTs deposited on the polished region. The SPF was placed inside a gas-tight metallic chamber isolated from the environment (Figure S1, Supporting Information). The precise control of composition of the gaseous mixtures of $NO_2$ with dry or wet air was made using mass flow controllers; the constant conditions of dynamic flow were maintained at 100 mL $min^{-1}$. By varying the gas composition within the cell, the sensor's response to nitrogen dioxide was investigated.

The optical and resistive responses of the sensor to $NO_2$ are shown in **Figures 2C**. The film thickness was approximately 25 nm. Here and in all the test the film covered the whole length of the side-polished region (3 mm). In this experiment, $NO_2$ and dry air were alternately introduced into the gas chamber: the analyte in the mixture with dry air was introduced for three minutes, followed by three minutes of dry air. These comparative tests were conducted using 100 ppm $NO_2$ in the mixture with air. Optical and electrical signals were recorded simultaneously during the tests.

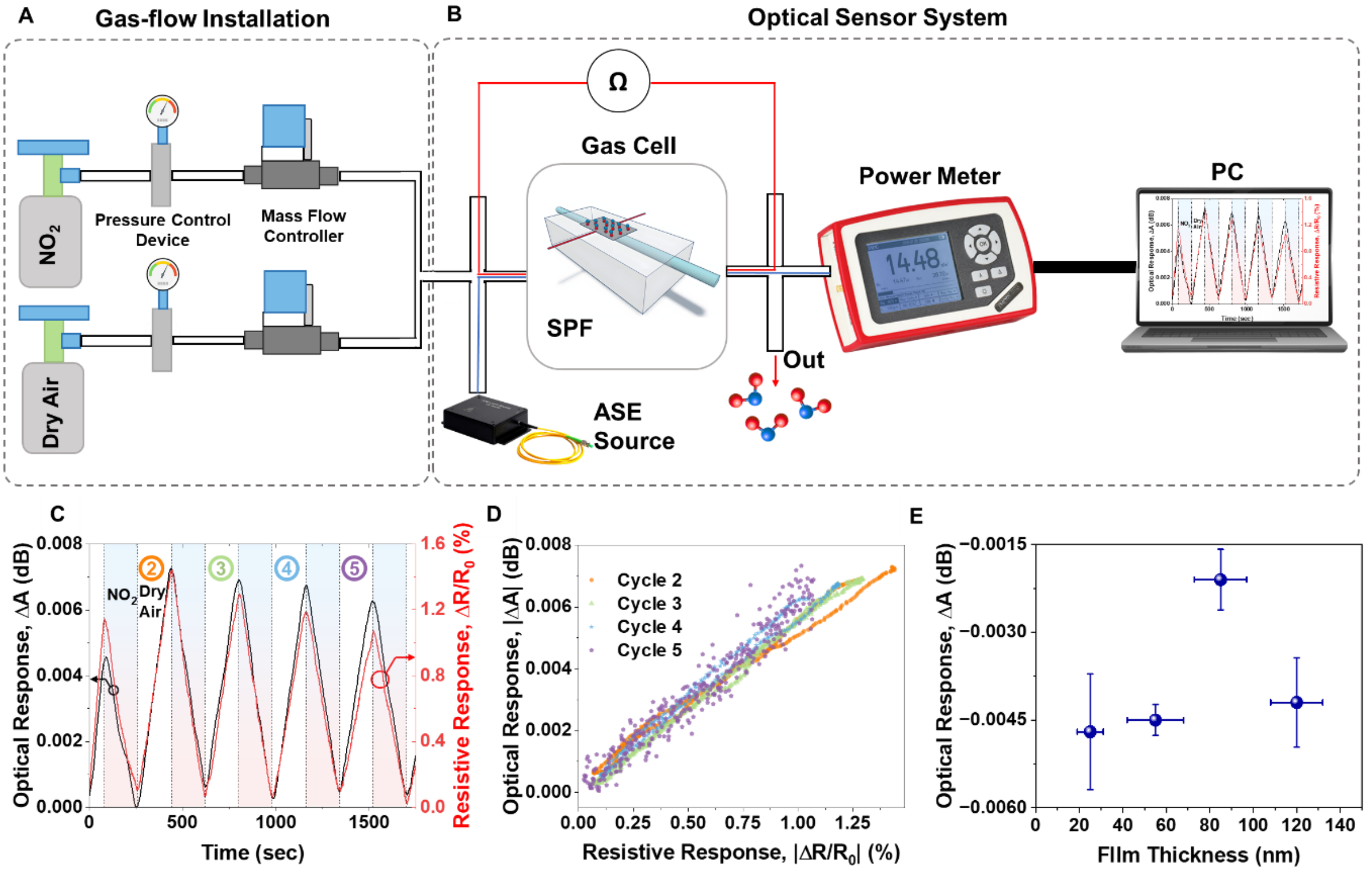

**Figure 2.** Schematic illustration of the setup for gas sensing performance tests: A) Gas-flow installation for obtaining specific composition of a $NO_2$ in the mixture with air, 100 mL $min^{-1}$. B) Optical sensor system. Light from ASE source propagates through SPF and is measured from the output by power meter connected to PC. C) Optical and resistive responses to 100 ppm analyte mixed with air. D) Correlation between optical and resistive responses. E) Dependence of the optical response on film thickness.

Both optical and resistive signals show a strong negative response to the analyte, demonstrating a decrease in the resistance and absorbance under gas exposure (**Figure 2C**). In this and all the following experiments, the colored sections indicate the purging of the analyte mixed with air through the chamber, while white ones depict dry air pulses. As the gases were switched multiple times, the resulting sensor readings formed a sawtooth pattern, with the slope change occurring shortly after the switching to pure air. Because the first gas exposure cycle differs significantly from the following and does not always represent the interaction mechanism, the response starting from the 2$^{nd}$ cycle will be given further.[52] **Figure 2D** illustrates the linear correlation between optical and resistive responses, confirming that the mechanisms behind optical and resistive interaction are the same for the studied SWCNT films - a modulation of concentration of charge carriers by hole doping, as discussed above. This confirms the finding of the previous work that for considered film thickness intertube barriers play a minor role in electrical sensing response of SWCNT films.[19,21]

Next, we investigated the dependence of the response on the film thickness (Figure 2E and Figure S2, Supporting Information for AFM measurements of the films). We found that the response is almost the same for all thicknesses, but, surprisingly, 85 nm is falling out towards smaller response. To understand the mechanism behind such unexpected behavior and motivated by the intrinsic optical anisotropy of SWCNTs[53], we measured polarization-dependent responses for SWCNT films of varying thicknesses. We used fiber polarization beam splitter (PBS) and polarization-maintaining (PM) side-polished optical fiber covered by SWCNT film. The light from ASE passed through PBS and was divided into two orthogonal polarizations: one that lies in the plane of the SWCNT film (TE); the other one is perpendicular to the film (TM). SEM image (see Figure S3, Supporting Information) shows that SWCNTs are mostly lying in the plane of the film in random orientation thus TE-polarization can be oriented both parallel and normal to the SWCNTs, while TM-polarized light is always orthogonal. The optical response for the polarized light in comparison with unpolarized is shown in **Figure 3**; the responses are presented in the same scale for better visibility of the effect which will be discussed shortly. For TM polarization, we always observe negative response as described previously; its amplitude grows almost 10 times (from -0.004 dB to -0.036 dB) as film thickness increases from 25 to 120 nm. However, for TE polarization we see strikingly different behavior.

For 25 nm and 55 nm films, we still observe a negative response, whose amplitude decreases for higher thickness. Finally, for 85 nm and 120 nm, it changes the sign to a positive response and reaches +0.02 dB for the 120 nm film. To describe this behavior, we refer to the previous investigation[54], where the nonlinear absorption of the SWCNT-covered waveguide was investigated and a similar effect for TE polarization was observed. We performed COMSOL Multiphysics simulations of absorbance dependence on various film thicknesses for both polarizations. To provide the best correspondence with experimental results, the real part of the refractive index of SWCNT film, $n_{CNT}$, was taken as 1.9 for TE polarization and 1.65 for TM polarization; imaginary parts of SWCNT refractive indices, $\varkappa$, were 0.9 and 0.45 for TE and TM-polarizations, respectively in agreement with measurements presented in ref. [53].

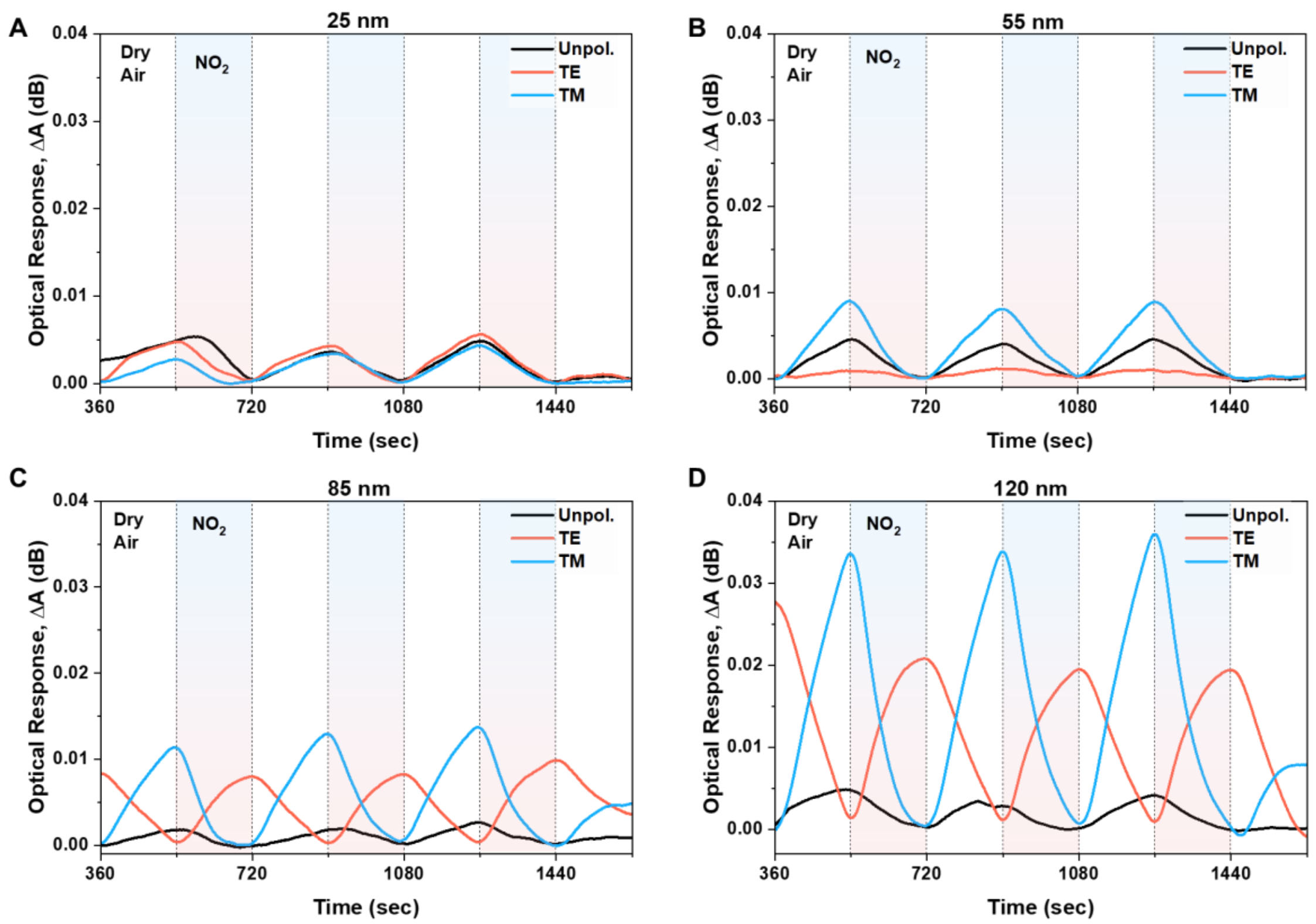

**Figure 3.** Results of optical response to $NO_2$, 100 ppm, in air depending on SWCNT film thickness and polarization for the films with thickness of A) 25, B) 55, C) 85 and D) 120 nm, accordingly.

We model the $NO_2$-induced Fermi-level shift as a small decrease in the imaginary part of the refractive index $\varkappa$. The change in the real part of refractive index, which also accompanies CNT doping, we consider as a smaller effect and neglect in our analysis.[54] The losses of light propagating through the SPF covered with SWCNT film ($\alpha_{SPF}$) can be described as[55]:

$$\alpha_{SPF} = \frac{4\pi}{\lambda}\frac{\iint_{-\infty}^{+\infty}\varkappa(x,y)|E(x,y)|^2dxdy}{\iint_{-\infty}^{+\infty}|E(x,y)|^2dxdy} = \frac{4\pi}{\lambda}\varkappa I, \quad (1)$$

$$I = \frac{\iint_{over\,film}|E(x,y)|^2dxdy}{\iint_{-\infty}^{+\infty}|E(x,y)|^2dxdy}, \quad (2)$$

where $I$ is the overlap integral of the mode profile with the carbon nanotube film and $\lambda$ is the wavelength of mode. Thus, the losses of mode depend not only on the imaginary part of SWCNT refractive index, but also on the overlap of the mode with SWCNT film. **Figure 4A** represents the distribution of the electric field for large (left side) and small (right side) values of $\varkappa$ for 120 nm film for the TE mode. We see that mode field distribution is asymmetric and for smaller $\varkappa$ it tends to move in the direction of the film, increasing overlap integral and losses $\alpha_{SPF}$. For the TM mode (**Figure 4D**), the same shift in mode distribution is also observed, but it is much smaller and does not affect losses significantly. To validate our model, we compare calculated and experimental losses depending on the thickness of the film which show good agreement for both TE (**Figure 4B**) and TM (**Figure 4E**) modes. Interestingly, TE polarization losses show nonmonotonic behavior, increasing at low film thicknesses and decreasing for thicknesses higher than 85 nm. This effect can be described by a reduction of the overlap integral with increase of the film thickness. The response to $NO_2$ is given by $\Delta A = \frac{\partial A}{\partial \varkappa}\Delta\varkappa$. Suggesting that for the same $NO_2$ concentration $\Delta\varkappa$ is the same for any film thickness, we plot calculated $-\frac{\partial A}{\partial \varkappa}$ (here minus comes because $\Delta\varkappa$ is negative under gas action) as a function of film thickness *(d)* and compare it to measured response $\Delta A$ (dB). In **Figure 4C** we see that our model can describe the behavior of TE mode, where we observe change of the sign of the response $\Delta A$ for thicker films. In brief, for thick films reduction of $\varkappa$ under gas exposure pulls the mode profile towards the film, increasing overlap integral and resulting absorption. The same model predicts always negative response for TM mode (**Figure 4D**) in agreement with the experiment. By comparing the $-\frac{\partial A}{\partial \varkappa}(d)$ with experimental data we can find $\Delta\varkappa \approx -0.006$ RIU for both polarizations, which provide a rough estimation of the gas action in concentration of 100 ppm in the mixture with air for the synthesized SWCNT films. This model also describes Figure 2E, where response for unpolarized light is plotted, see Figure S4, Supporting Information.

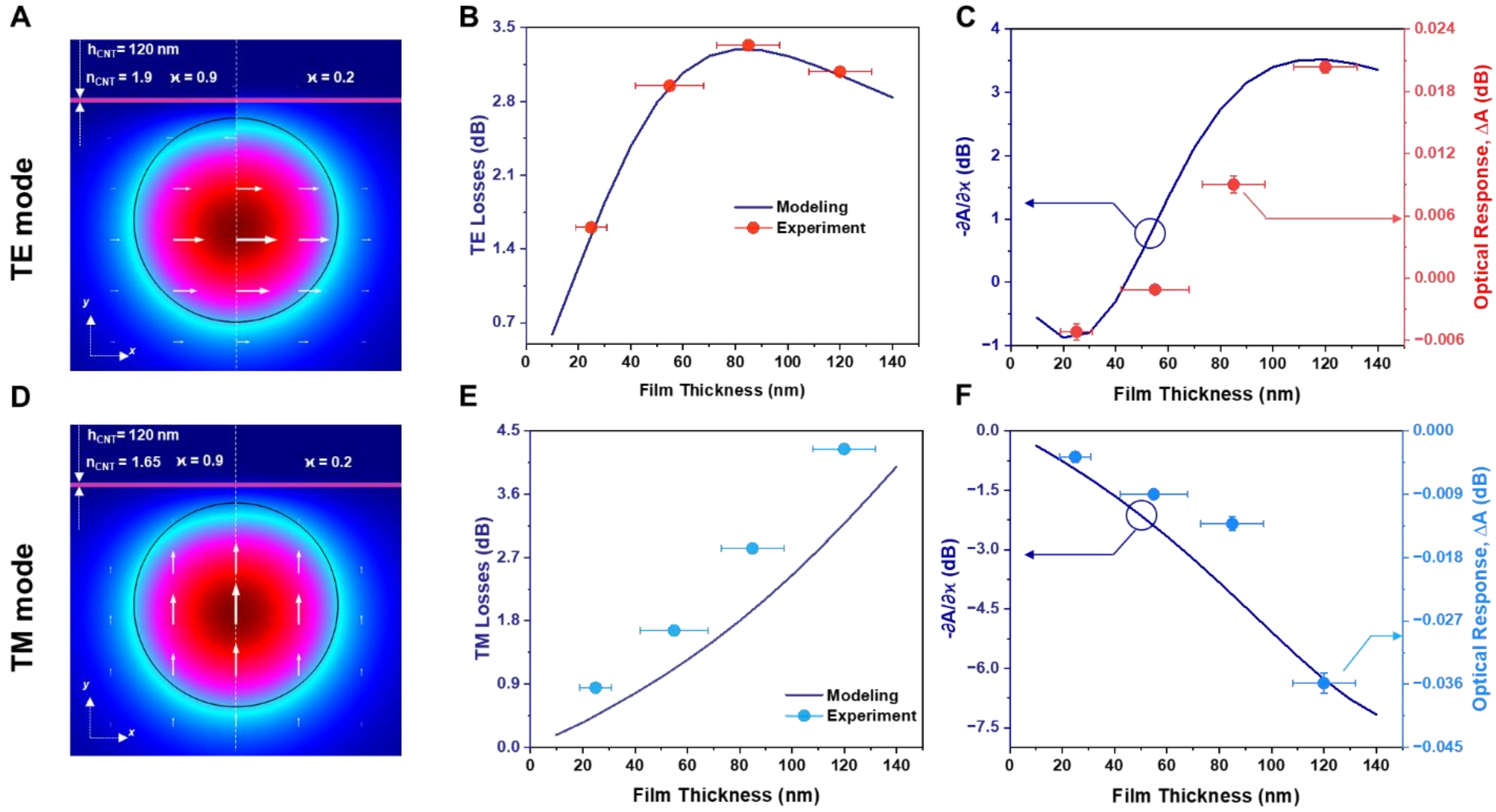

**Figure 4.** A) Results of the simulation of the TE field distribution in the SPF-SWCNT. B) Experimental (red) and modeled (dark blue) dependence of the attenuation for TE polarization. C) Derivative of absorbance as a function of film thickness for TE-polarized light. D-F) replicate A-C) for TM-polarized light, respectively.

Following our analysis, we used a 120 nm film thickness and TM-polarized light, which provided the highest response, to evaluate the LOD and sensitivity. For this we introduced analyte concentrations from 1 to 100 ppm in the mixture with air; exposure to analyte lasted 6 minutes and pure dry air pulse – 9 minutes (**Figure S5A**, Supporting Information). Exposure to each concentration of $NO_2$ was repeated 3 times, followed by 30-min relaxation period for desorption. The obtained calibration curve (**Figure 5A**) demonstrates linear dependence of the response to analyte followed by a saturation plateau, which is typical for $NO_2$.[56]

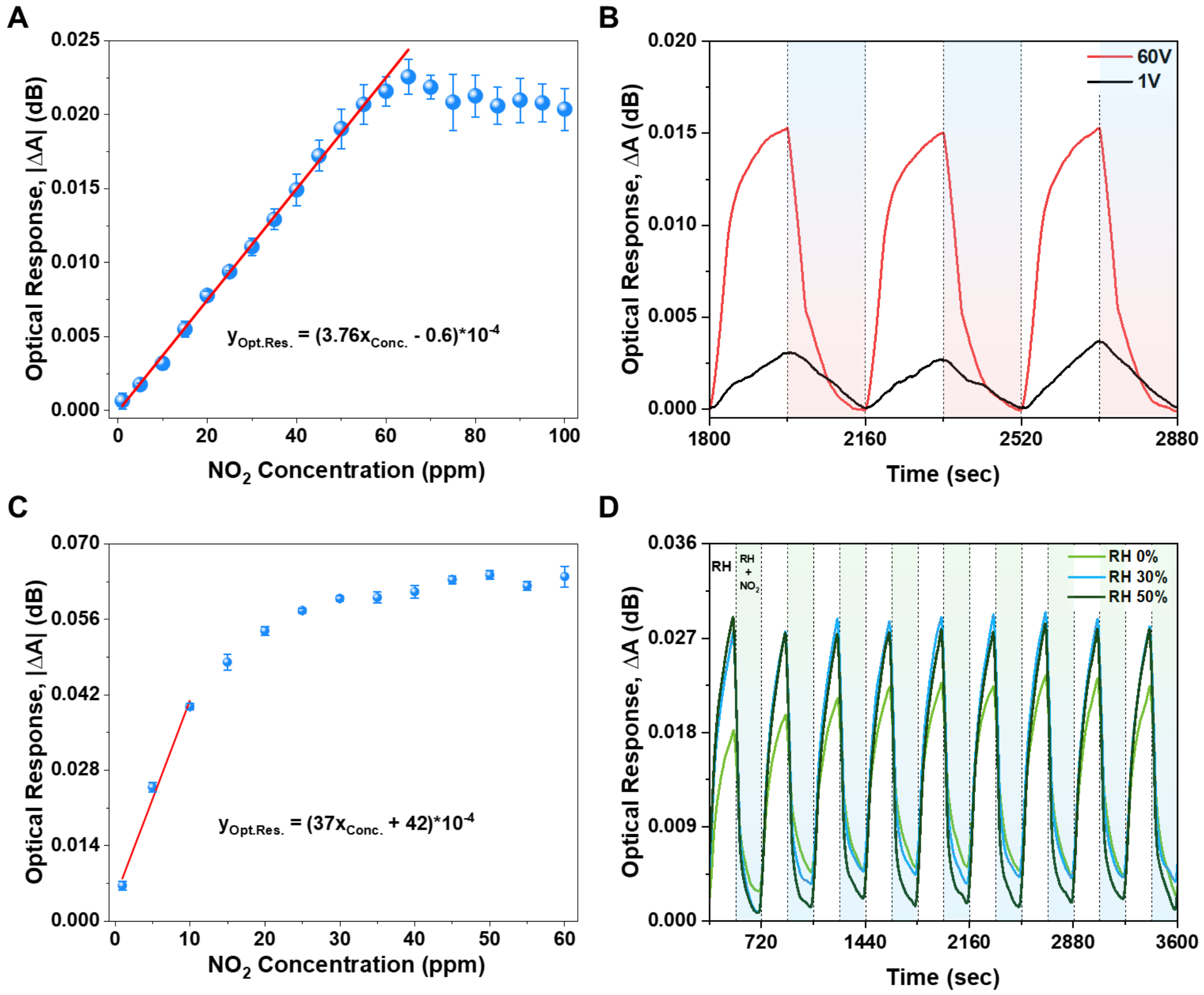

**Figure 5.** A) Calibration curve for TM-polarized light and 120 nm thick SWCNT film in the linear range. B) Optical response by short-circuit annealing. C) Calibration curve for TM-polarized light and 120 nm thick SWCNT film in the linear range with heating. D) Optical Response under $NO_2$ exposure at different relative humidity values.

Using the experimental data, sensor's LOD and sensitivity were calculated[21]:

$$LoD = 3.3 \cdot \frac{\sigma}{b} = 1.37\ ppm, \tag{3}$$

$$S_O = \frac{\Delta A}{\Delta C} = 0.00035\ dB\ ppm^{-1}, \tag{4}$$

where $\sigma$ – standard deviation of the sensor's signal, $b$ – slope of the calibration curve, $S_o$– optical sensitivity, respectively, $A$ – absorbance, $C$ – concentration.

Next, we studied how temperature affects sensor response and recovery kinetics; this investigation is important because heating facilitates desorption process and many nanomaterial-based sensors are working at elevated temperatures[13–16] (**Figure 5B**, Figure S6B,

Supporting Information). The conductivity of carbon nanotubes allows them to be heated just by passing a current up to damage threshold, which is around 400 °C in presence of oxygen.[57] To find the dependence of temperature of the SWCNT film *versus* the applied voltage (passing current), we employed temperature dependence of G-peak position on the Raman spectra (see **Figure S7** in Supporting Information).[50] The comparison of responses under 1 V and 60 V (120 °C) for 50 ppm $NO_2$ is shown in **Figure 5B**. The response increased by almost a factor of five for the same 3 minutes exposure to analyte. Additionally, at higher temperatures, the drift is less prominent, suggesting an optimal operational temperature regime. We limit the temperature to 130 °C due to formation of nitrate forms[58] which prevent analyte desorption when SWCNTs are exposed to $NO_2$ at temperatures exceeding 150 °C.[58] With annealing, 120 nm thick film and TM-polarized light (**Figure 5C**), at these conditions the estimated LOD and sensitivity are 400 ppb and 0.0037 dB $ppm^{-1}$, respectively. These results correspond to typical nanomaterial-based $NO_2$ optical sensors, summarized in **Table 1**.

**Table 1.** Comparison of gas sensing performance with optical fiber sensors.

| Materials | Platform | Operating Wavelength (nm) | Operating Temperature | Sensitivity | LOD | Detection Range | Selectivity | Response Time | Ref. |
|---|---|---|---|---|---|---|---|---|---|
| $MoWS_2$ | SPF | NA | RT | 86% ($NH_3$)<br>72% ($NO_2$)<br>48% (CO) | NA | 0-500 ppm | $CH_4$, $CH_2O$, $NH_3$, CO, $N1O_2$ | 21 sec<br>42 sec<br>93 sec | 59 |
| 2D $SnS_2$ | SPF | 473 | RT | 84.3 uW/ppm | 464 ppt | 0-50 ppb | $H_2$, $H_2S$, CO, $CH_4$, $CO_2$ | 7.7 min | 60 |
| 2D $WO_x$ | SPF | 1550 | 160 °C | 43.18 uW/ppm | 8 ppb | 0-704 ppb | $H_2$, $H_2S$, $CH_4$, $CO_2$ | 11 min | 29 |
| RGO | FBG | 1545 | RT | 11 pm/ppm | 500 ppb | 0.5 – 3 ppm | CO, $CO_2$ | 12.3 min | 61 |
| SFA/DMNA | Doped optical fiber core | 420, 520 | RT | 14.5 ab/ppm | 193 ppb | 0.193 – 0.24 ppm | NA | 8 min | 62 |
| Silica-$LuPC_2$ | Tip end of singlemode fiber | 660, 365 | RT | 0.05 uW/ppm | 50 ppb | 0.7 – 7 ppm | NA | 6 min | 63 |
| SWCNTs | SPF | 1550 | RT | 0.0004 db/ppm | 1.4 ppm | 0 – 100 ppm | $NH_3$ | NA | This work |
| SWCNTs | SPF | 1550 | 130 °C | 0.004 db/ppm | 400 ppb | 0 – 100 ppm | $NH_3$ | 8.6 min | This work |

Finally, for room temperature sensor it is important to verify that it can operate at humid conditions. The relative humidity was set to 0%, 30% and 50%; concentration of analyte was set to 50 ppm in the mixture with air. **Figure 5D** shows a clear response when $NO_2$ was introduced into the gas chamber. Moreover, the signal amplitude increases in humid conditions. It can be attributed to analyte reaction with surface adsorbing water which creates nitrous and nitric acids providing effective p-doping.[51]

## 3. Conclusion

In summary, we have developed and rigorously characterized a high-performance optical gas sensor by integrating SWCNT thin film with side-polished optical fiber. This study demonstrates that the performance of evanescent wave devices is dictated not only by the intrinsic material sensitivity (such as the change in complex refractive index per ppm, ΔRI/ppm) but also by the transduction efficiency arising from perturbations of the waveguide's evanescent field. A pronounced polarization- and thickness-dependent response was observed: for TE polarization, the signal magnitude decreases with increasing SWCNT thickness, vanishes at around 60 nm, and then reverses sign, whereas for TM polarization the response increases monotonically with thickness. Numerical modeling attributes this non-monotonic behavior to gas-induced mode reshaping: changes in the SWCNT optical constants selectively modify the overlap integral and, as a result, propagation loss of different guided modes. Based on this insight, optimal operating conditions were identified—a 120 nm film interrogated with TM-polarized light at 1550 nm—yielding an optical limit of detection of 400 ppb, a sensitivity of 0.0037 dB $ppm^{-1}$, and response/recovery times of 8.6 and 18.5 minutes, respectively, while preserving stable performance under humid conditions. These findings not only validate the SWCNT–SPF architecture as a practical platform for industrial and environmental gas monitoring but, more fundamentally, reveal mode reshaping mechanism as an important determinant of sensor response which should be accounted for the in design and optimization of next-generation fiber-optic and integrated photonic sensors reliant on evanescent-field interactions.

## 4. Experimental Section/Methods

*Synthesis of SWCNT:* Single-walled carbon nanotubes were synthesized using the aerosol chemical vapor deposition method based on the Boudouard reaction, with ferrocene serving as the catalyst precursor and carbon monoxide as the carbon source.[45] A flow of CO and ferrocene vapor, heated to 880 °C in a tubular furnace reactor, resulted in the formation of iron-based catalyst nanoparticles and subsequently SWCNTs.[64] The SWCNT films were collected downstream of the reactor on a nitrocellulose filter. These films consisted of a mixture of semiconducting and metallic SWCNTs arranged in a random network with a predominant in-plane orientation. The thickness of the SWCNT film on the filter was precisely controlled by varying the collection time.

*Device Fabrication:* The device was fabricated using an optical fiber with polished cladding embedded in a quartz block. The outputs of the fiber-optic block were connected to optical fibers using epoxy resin. A thin film of SWCNTs was deposited directly onto the side-polished fiber using a dry-transfer technique, forming the sensing element. To enable electrical resistance measurements during gas adsorption and desorption, two metallic wires were affixed to the SWCNT film using silver conductive epoxy adhesive. These wires served as electrodes for monitoring the resistive response of the SWCNTs.

**Acknowledgements**

The authors thank Dr. Ignat Rakov for the aerosol CVD reactor visualization. This work was supported by the Russian Science Foundation Grant No. 22-13-00436-П.

**Data Availability Statement**

The data that support the findings of this study are available from the corresponding author upon reasonable request.

# Supporting Information

## Harnessing Evanescent Wave Interaction for Enhanced Optical $NO_2$ Detection with Carbon Nanotube-Coated Side-Polished Fiber

*Egor O. Zhermolenko, Khasan A. Akhmadiev, Aram A. Mkrtchyan, Fedor S. Fedorov, Anastasiia S. Netrusova, Aliya R. Vildanova, Dmitry V. Krasnikov, Albert G. Nasibulin and Yuriy G. Gladush**

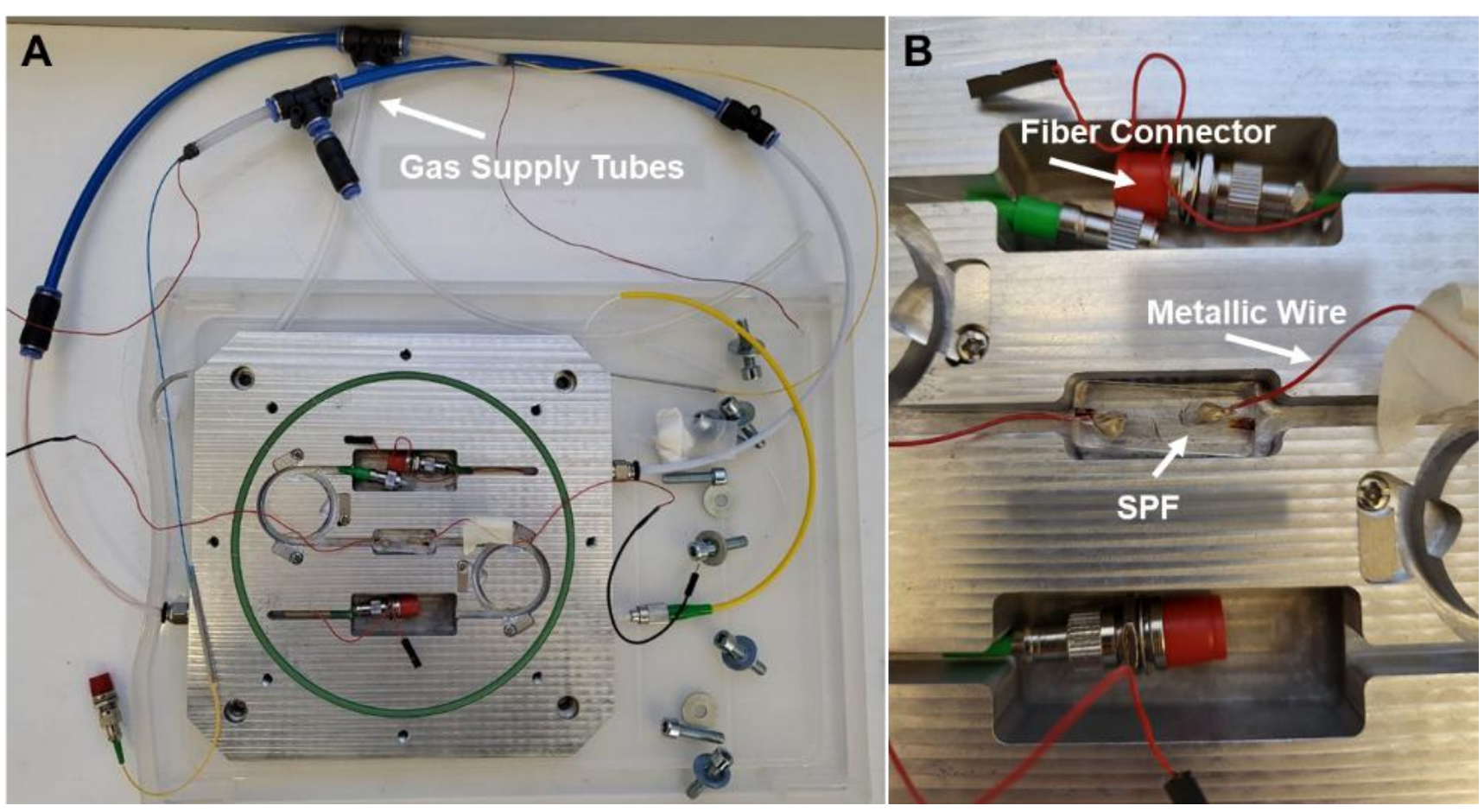

**Figure S1.** A) Gas Cell Close-Up. B) SPF positioned in gas chamber.

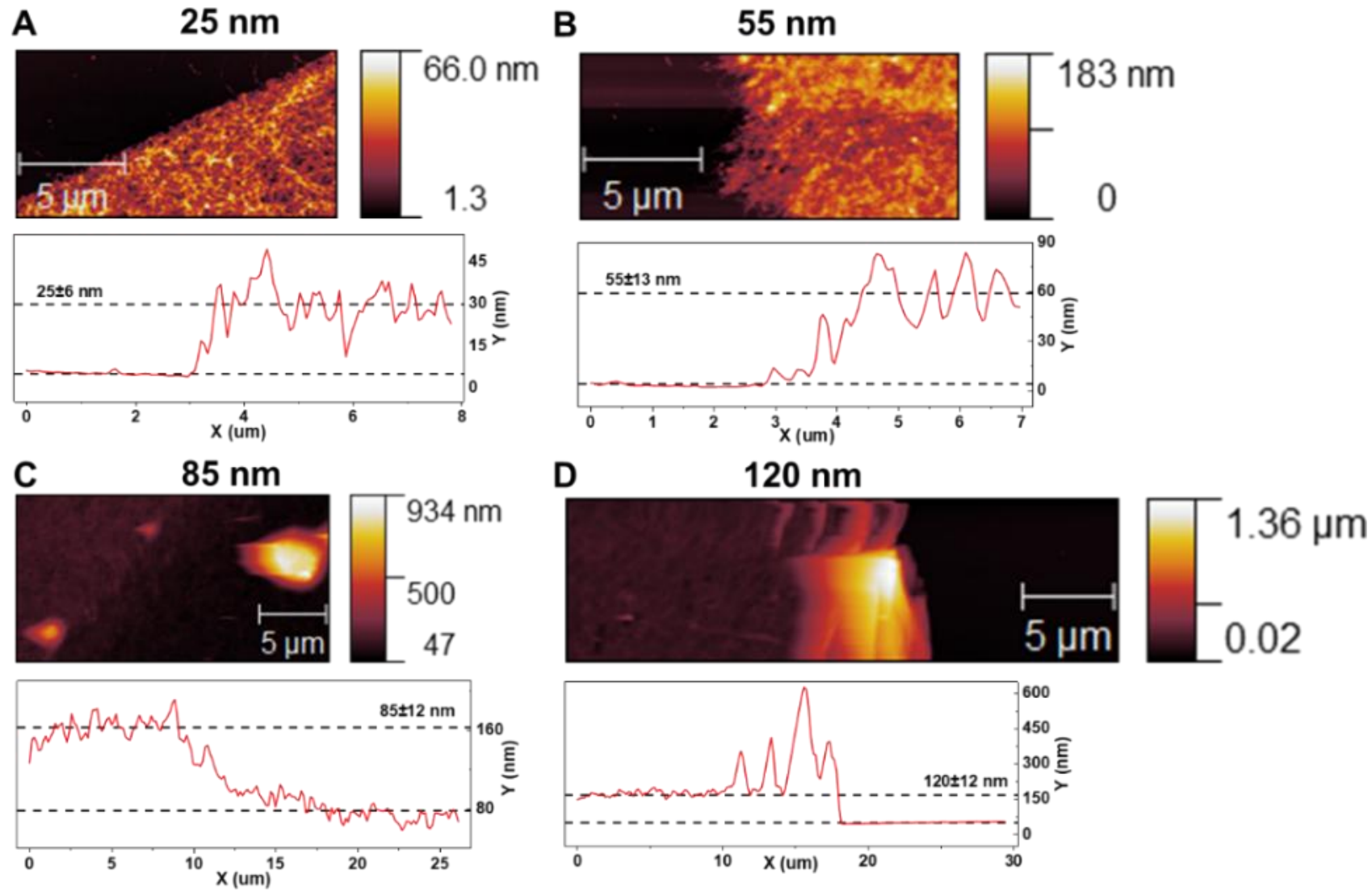

**Figure S2.** AFM images of SWCNT films with thicknesses of A) 25, B) 55, C) 85 and D) 120 nm, accordingly.

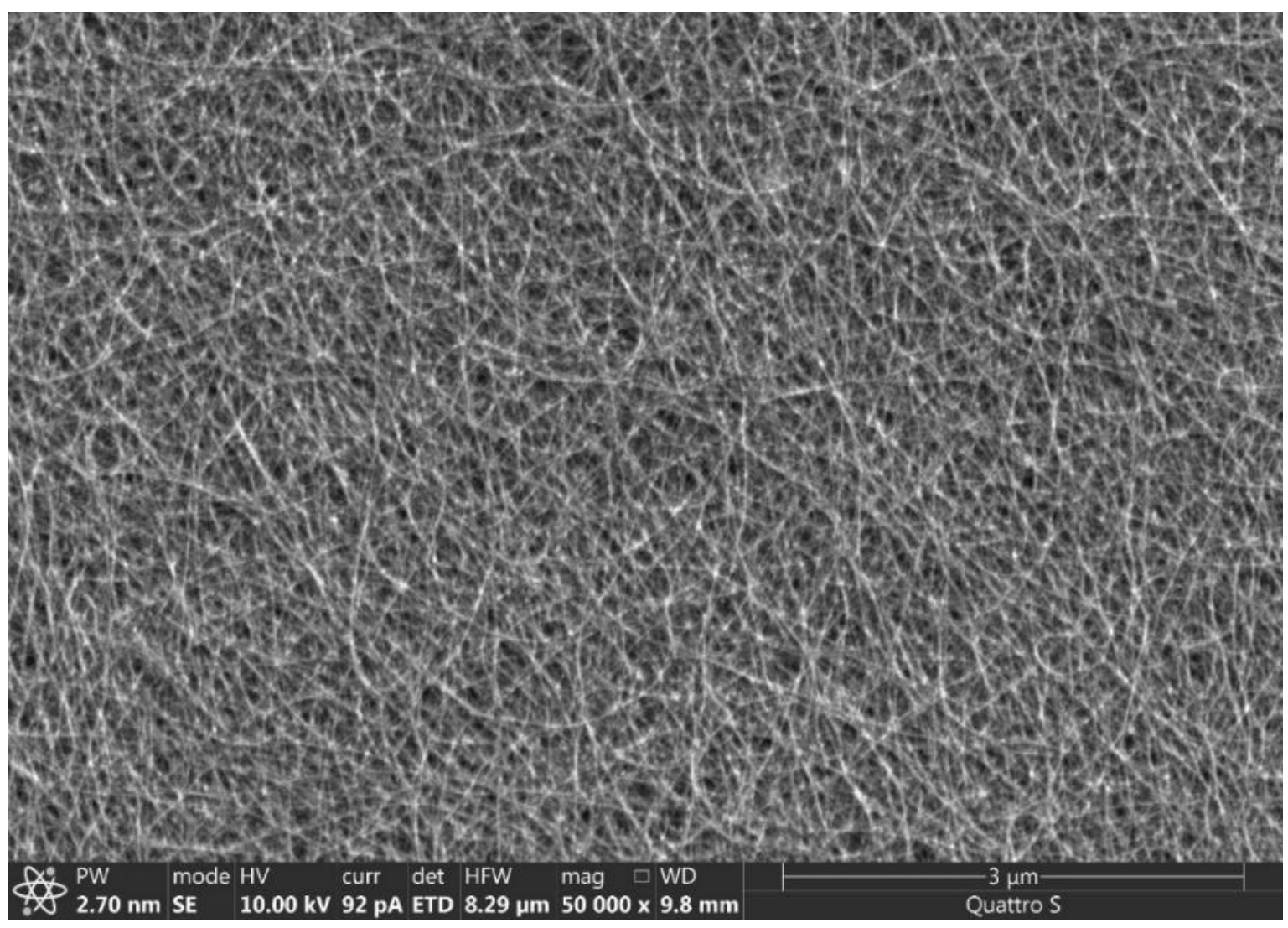

**Figure S3.** SEM image of SWCNT film.

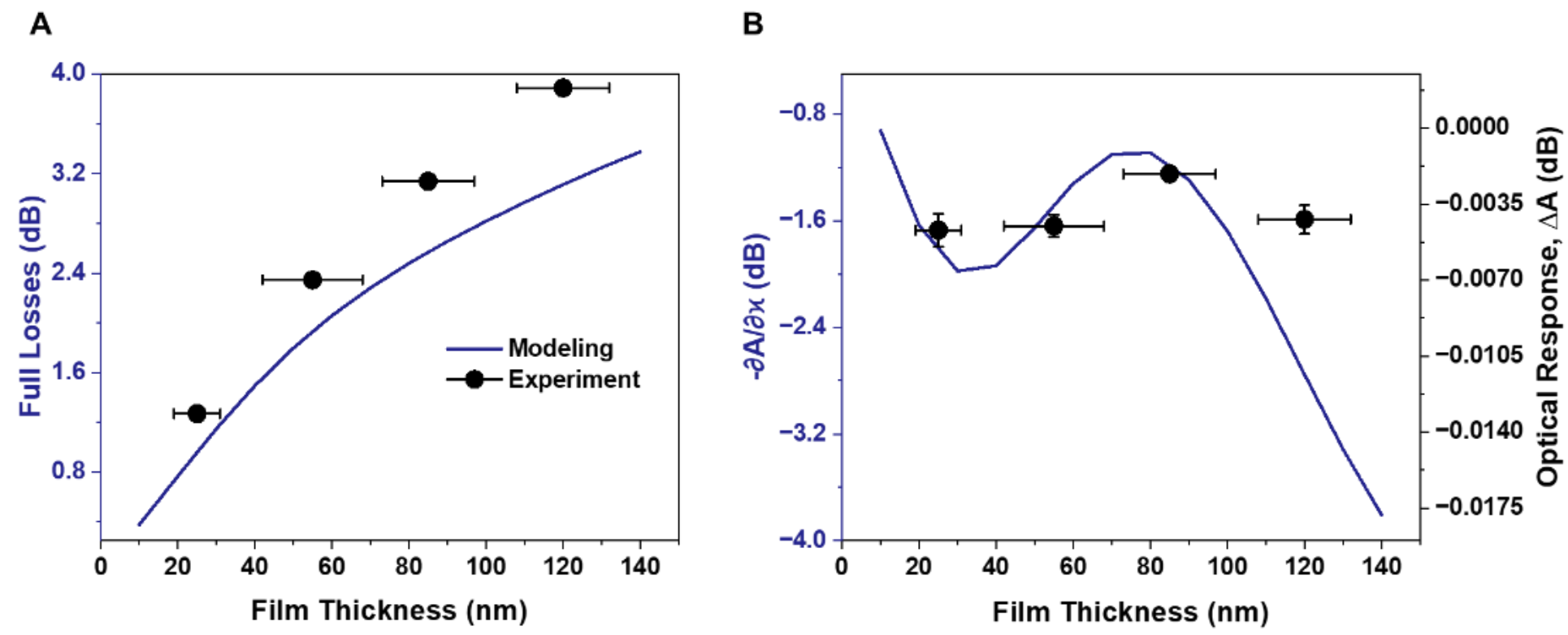

**Figure S4.** A) Experimental (black) and modeled (dark blue) dependence of the attenuation for unpolarized light. B) Derivative of absorbance as a function of film thickness for unpolarized light.

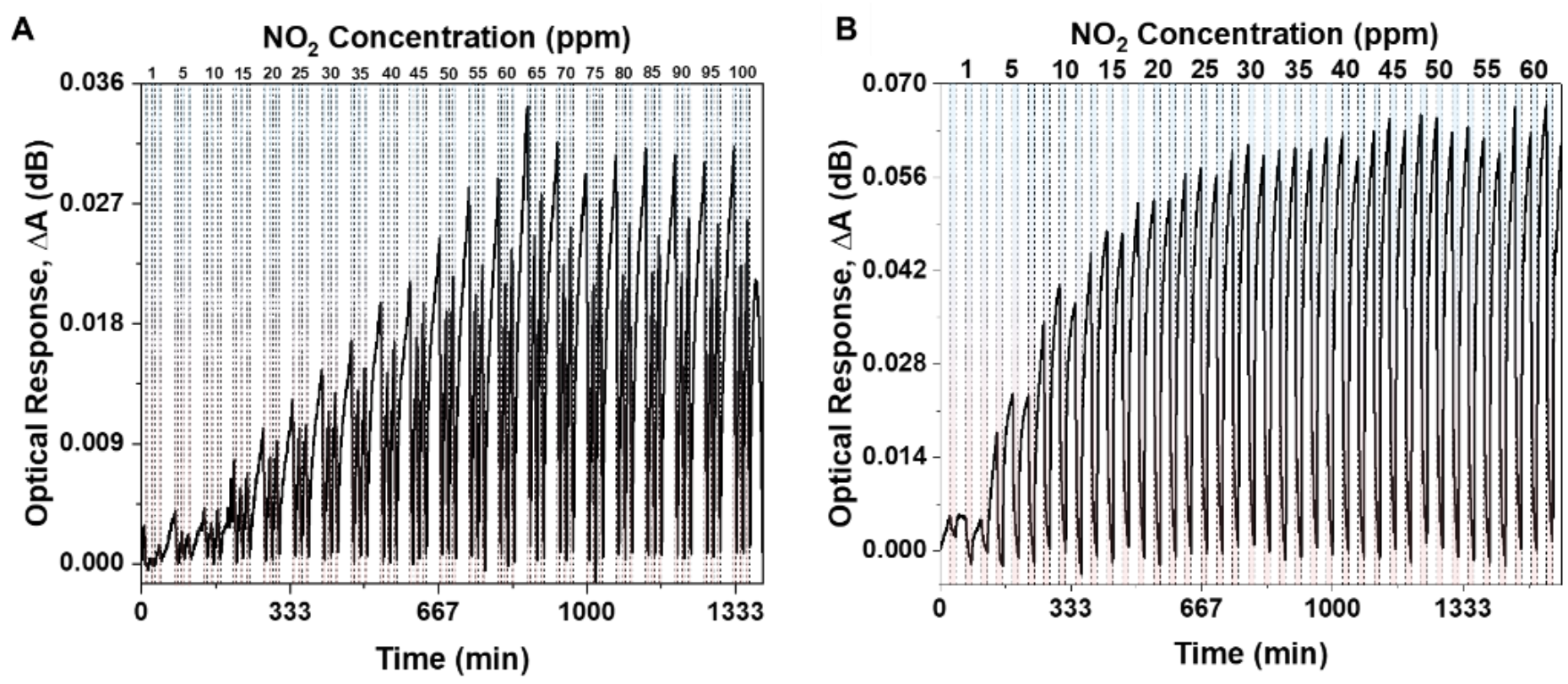

**Figure S5.** Limit of detection estimation using TM-polarized light and 120 nm thick SWCNT film. A) Without annealing. B) With annealing (130 ºC).

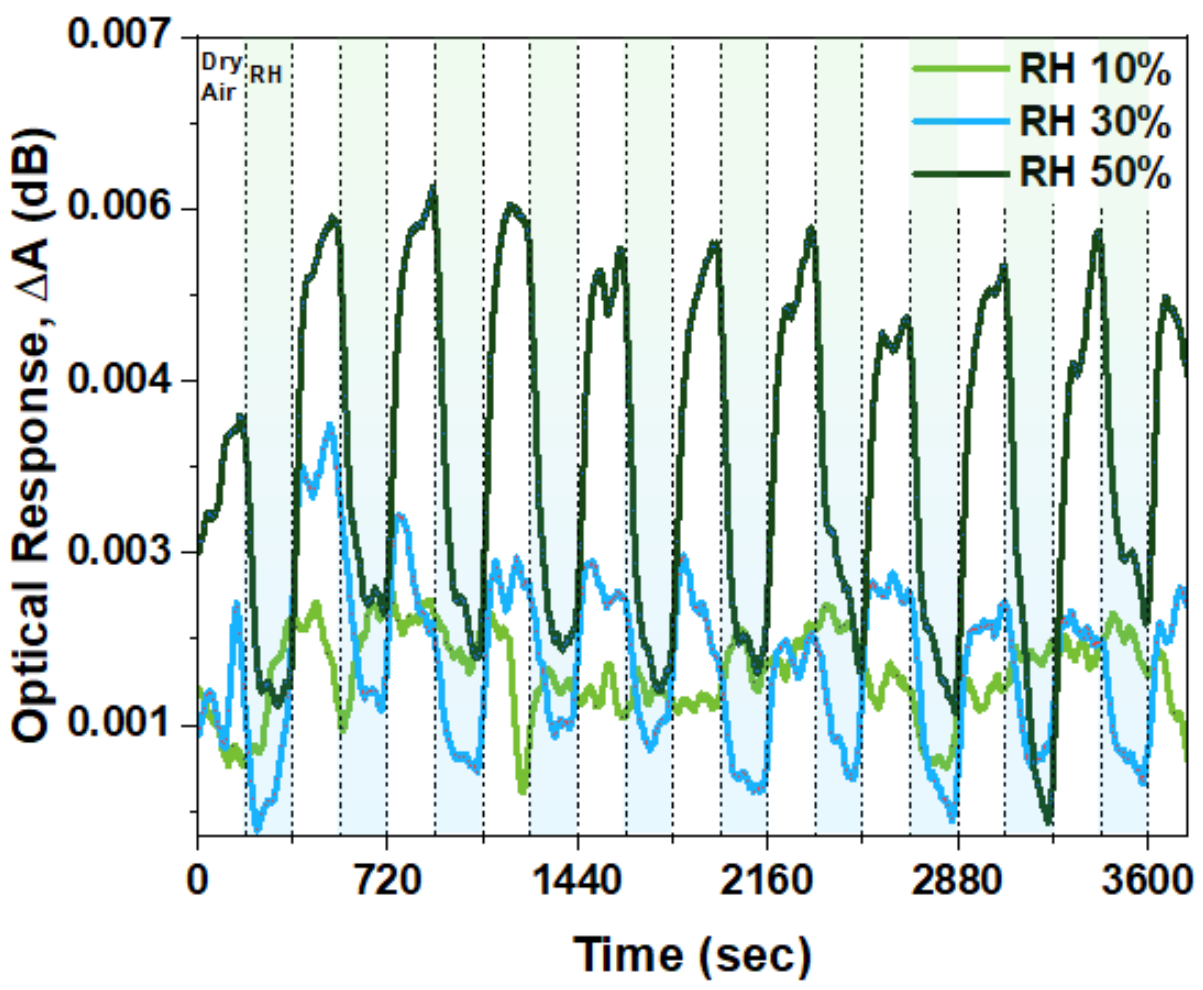

**Figure S6**. Optical response to humidity.

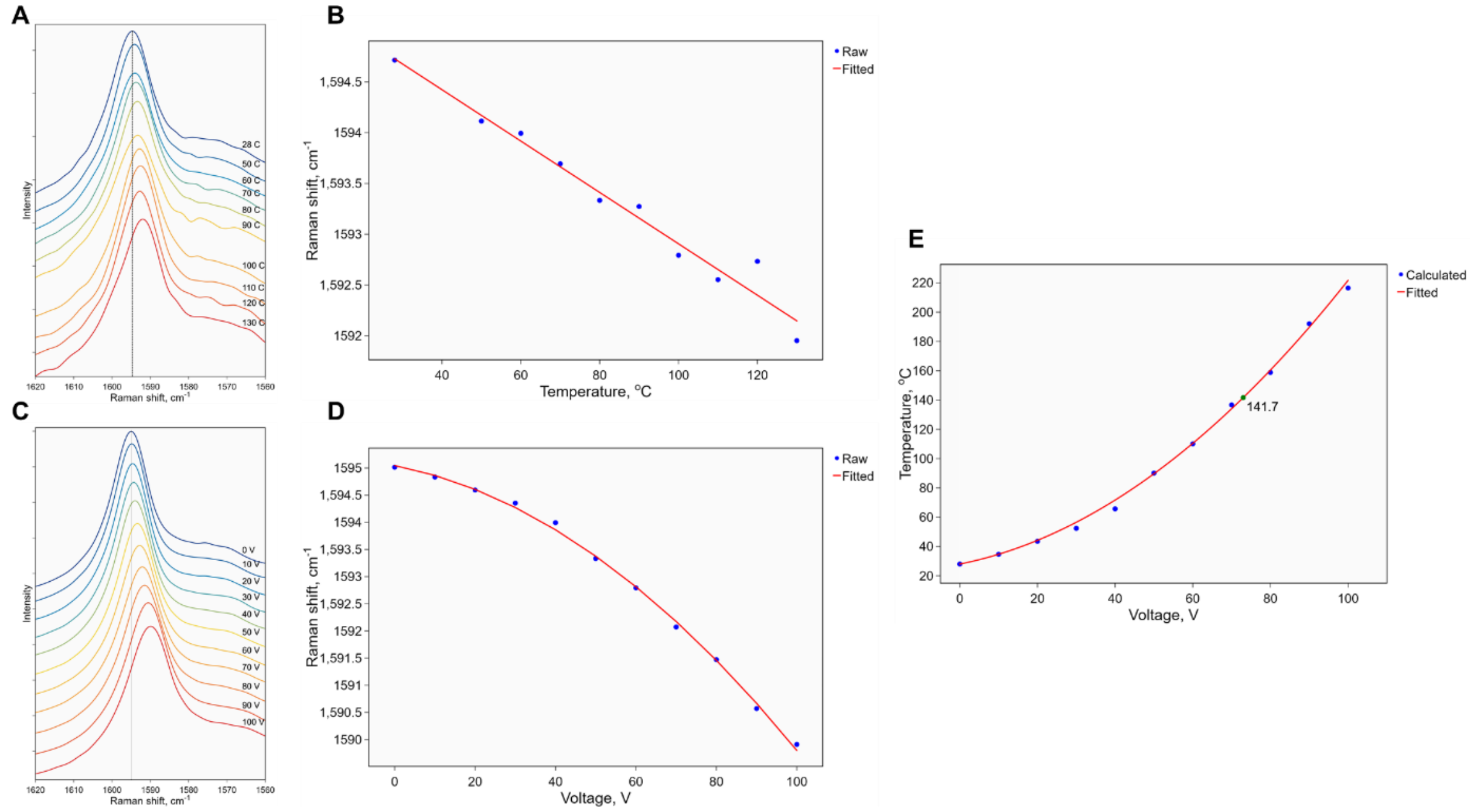

**Figure S7**. A) Raman spectra of SWCNT G-band intensities with different temperatures. B) Calibration curve of Raman shift toward changes with temperature. C) Dependence of temperature over applied voltage. D) Raman spectra of SWCNT G-band intensities with different applied voltages. E) Calibration curve of Raman shift toward changes with applied voltage. F) Optical Signal Restoration using electrical annealing.

During our experiments we experienced signal deterioration after a prolonged exposure to analyte and environment due to chemisorption. Desorption of the residual analyte molecules can be achieved via electrical annealing. The analysis of Raman spectra shifts is conducted to determine the relationship between SWCNT film's temperature and applied voltage. **Figure**

**S7A** shows the Raman shifts of G-band peak with increasing environment temperature. As it can be seen, the dependence is linear, and the G-band shift coefficient is -0.027 $cm^{-1}$ $K^{-1}$ (**Figure S7B**). The Raman measurements were conducted while applying voltage are conducted and shift coefficient is used for calculating the material's temperature. **Figure S7C** shows the G-band peak shifts with increasing applied voltage. The Raman shift behavior is a quadratic function of voltage, and obtained coefficient is further utilized (**Figure S7D**). **Figure S7E** depicts the dependence of temperature on applied voltage. The required temperature for heating SWCNTs is about 140°C; therefore, the applied voltage should be lower than 75 V. To show sensor efficiency during annealing of SWCNT film, experiments from 1 V to 75 V of applied voltage are conducted with 50 ppm of $NO_2$. The optimal applied voltage to achieve the stable signal and low recovery time turned out to be 60 V.